\documentclass[pdftex,sn-mathphys-num,14pt]{sn-jnl}

\usepackage{balance}
\usepackage{graphicx}
\usepackage{color}
\usepackage{array}
\usepackage{multirow, makecell}
\usepackage{rotating}   
\usepackage{lscape}
\usepackage{url}
\usepackage{enumitem}
\usepackage{lmodern}
\usepackage{amsmath}

\begin{document}

\title[]{\large \textbf{Reddit Community Signals as a Supplementary Domain-Reputation Feature in Spam Classification: A Conservative Controlled Evaluation}}

\author*[]{\fnm{Fatima} \sur{Salcedo}}
\email{fatimiz49@gmail.com}

\author[]{\fnm{Jide} \sur{Edu}}
\email{jide.edu@strath.ac.uk}

\affil[]{%
  \orgdiv{Department of Computer and Information Science},
  \orgname{Faculty of Sciences, University of Strathclyde},
    \country{UK}
  }

\abstract{
Email spam and phishing remain persistent threats, and purely algorithmic classifiers may struggle with unfamiliar senders, evolving message content, and cold-start domains. This study evaluates whether Reddit-derived community signals can serve as a supplementary domain-reputation feature in email spam classification. We use a preprocessed SpamAssassin-derived corpus of 5,582 emails and compare three feature configurations: (1) a baseline using TF-IDF and structural features; (2) a traditional-enhanced configuration adding urgency and action scores; and (3) a Reddit-enhanced configuration adding a binary discussion-presence indicator, a keyword-based Reddit score, and an urgency–Reddit interaction term. We train Logistic Regression, Random Forest, and Support Vector Machine models. In the preliminary single-split evaluation, the Random Forest model with Reddit-enhanced features achieves 97.94\% accuracy, a 0.55 percentage-point improvement over the baseline, although the improvement is not statistically significant in the primary test split. Logistic Regression performance declines when additional features are introduced, and the SVM performs poorly and is treated as a negative control. Reddit signals appear most informative for rare domains, but approximately 70\% of emails receive no Reddit-derived signal. Because the Reddit data are temporally later than the email corpus, we do not interpret the results as evidence of real-time detection. Instead, the study provides a conservative evaluation of community-derived signals as a supplementary, asynchronous domain-reputation feature and identifies the additional validation required before production deployment, including domain-grouped validation, temporally aligned contemporary data, and adversarial robustness testing.
}

\keywords{\sep spam detection, \sep email security, \sep domain reputation, \sep community intelligence, \sep Reddit, \sep threat intelligence, \sep feature engineering, \sep machine learning validation, \sep temporal validity, \sep auxiliary filtering}

\maketitle

\section{Introduction}

Email spam and phishing continue to pose significant cybersecurity risks~\cite{ORO24}. Despite decades of improvement in filtering technology, attackers regularly adapt their messages to evade purely algorithmic defenses. Modern phishing emails often imitate legitimate organizational communication, use grammatically correct language, and incorporate credible branding, making content-only classification increasingly difficult~\cite{Goo07,Chr14}. Empirical field studies also indicate that a non-trivial proportion of users remain susceptible to phishing, suggesting that existing filters do not eliminate user-facing risk~\cite{Som19}. This persistence suggests that spam detection systems may benefit from signals that go beyond the text of the email itself.

One promising source of supplementary information is online community discussion. Platforms such as Reddit host user-generated discussions about suspicious domains, scams, phishing attempts, and malicious senders~\cite{Mur23,Red09}. In security-focused communities, users often report newly observed threats, share experiences with suspicious links, and collectively assess whether a domain or sender appears malicious~\cite{Gle18}. Because such discussions can arise before formal reputation systems accumulate sufficient historical data, they may be particularly useful for unfamiliar or low-trust domains.

Reddit offers a unique opportunity for threat intelligence due to its structural characteristics, particularly its upvote/downvote mechanism, which serves as a natural filtering system that highlights the most relevant and accurate information through community consensus. In addition, specialized subreddits (e.g., r/Scams, r/phishing) serve as repositories of collective knowledge about malicious actors. Furthermore, the temporal persistence of discussions allows domains active years ago to leave traces in community memory, providing historical context that traditional reputation systems might miss.

However, community signals are not a panacea. Reddit is not a bot-free environment, and community discussions can be manipulated through astroturfing, coordinated voting, disinformation campaigns, or bad-faith participation~\cite{Gle18, Mur23}. In addition, community attention is sparse: many domains are never discussed, and domains that are discussed may be discussed for reasons unrelated to email spam, such as technical support, product complaints, or general security news. Finally, the temporal alignment between email messages and community discussions is critical. Using current Reddit discussions to classify historical emails is only appropriate if the study is framed as an offline evaluation of domain-level reputation features, not as evidence that the system would have operated in real time.

This paper addresses the following question:

\begin{quote}
\textbf{Do Reddit-derived, domain-level community signals improve spam classification when treated as a supplementary domain-reputation feature, and under what conditions is their benefit most plausible?}
\end{quote}


The theoretical motivation draws on two bodies of work. First, the wisdom-of-crowds literature suggests that aggregated group judgments can be reliable under conditions of diversity, independence, decentralization, and appropriate aggregation~\cite{Sur04}. Second, ensemble learning research suggests that combining diverse signals or learners can improve robustness relative to single-feature or single-model systems~\cite{Die00}. We use these ideas as motivation, but we do not treat them as proof that Reddit-derived signals are automatically valid.

To assess the contribution of Reddit community-derived signals, we implement a three-stage feature-ablation design. The baseline configuration uses TF-IDF representations of email subject and body text together with structural features such as URL count and message length. The traditional-enhanced configuration adds urgency and action scores that quantify linguistic indicators commonly associated with social-engineering emails. The Reddit-enhanced configuration adds three community-derived features: a binary indicator of whether the sender domain has associated Reddit discussion, a continuous Reddit score derived from weighted keyword analysis of domain-specific discussions, and an interaction term between urgency and Reddit score.
%
We employ three complementary machine learning models: Logistic Regression (for interpretability), Random Forest (for capturing feature interactions), and Support Vector Machines (SVM) (for geometric validation). The models were selected for their ability to handle the dual imbalance challenges in the data (30-70 spam-ham distribution) and the variability of Reddit discussions. This multi-model approach ensures reliable conclusions about the value of community intelligence, irrespective of algorithmic choice.



The primary contributions of this work are as follows:
\begin{enumerate}[label=\arabic*., leftmargin=*]
\item \textbf{Theoretical Integration:} We bridge the disciplinary gap between human-computer interaction and machine learning in cybersecurity, demonstrating how community-sourced intelligence can augment and refine algorithmic detection approaches.
\item \textbf{Practical Framework:} We establish a reproducible framework for extracting and integrating Reddit-based social signals into existing spam detection pipelines, enabling empirical validation of crowd-derived threat data.
\item \textbf{Societal Implication:} We underscore the critical role of human collective intelligence in an era of automation, positing that online communities serve as essential defensive layers against threats emerging from AI-driven digital transformation.
\item \textbf{Practical deployment guidance.} We argue that the most defensible role for Reddit-derived intelligence is not as a primary real-time filter, but as an asynchronous, cached domain-reputation layer targeted primarily at unfamiliar or high-risk senders.
\end{enumerate}

\section{Literature Review}

The ongoing evolution of email-based threats has exposed critical vulnerabilities in conventional spam detection mechanisms, necessitating approaches that go beyond purely algorithmic solutions. While machine learning (ML) models have established robust baselines for classification, emerging adversarial tactics frequently outpace static model training, creating an adaptability gap in real-world deployments \cite{ORO24}. This review synthesizes current research to examine the theoretical and empirical potential of integrating community-driven intelligence—specifically from platforms like Reddit—into established frameworks. By analyzing the convergence of traditional detection methods, content-based analysis, collective wisdom theories, and reputation systems, we elucidate how human-derived signals can complement algorithmic processing to enhance classification accuracy and resilience against novel threats.

\subsection{Evolution of Machine Learning Approaches in Spam Detection}
The landscape of spam email detection has evolved from simple statistical models to complex ensemble methods, each contributing unique strengths to the classification challenge. Establishing robust baseline models remains critical for rigorous comparative analysis; however, their suitability depends on specific data characteristics and computational constraints \cite{Met06}.

\subsubsection{Probabilistic and Instance-Based Baselines}
The Naive Bayes classifier has historically served as a fundamental baseline due to its probabilistic interpretations of feature contributions. While useful for comparing performance against more sophisticated models, Naive Bayes relies on the assumption of feature independence and performs suboptimally when class distributions are imbalanced \cite{Jas03}. Given the significant imbalance between ham and spam emails in contemporary datasets, reliance solely on probabilistic assumptions is often limited. Similarly, K-Nearest Neighbour (KNN) offers an effective framework for understanding clustering patterns but was excluded from modern high-dimensional applications due to computational limitations. As an instance-based algorithm, KNN requires storing the entire training dataset and computing distances during inference \cite{Zha16}. This approach becomes computationally intensive when combining TF-IDF vectors with derived signals, presenting a bottleneck for scalable deployment.

\subsubsection{Ensemble Methods and Non-Linear Modeling}
Decision trees offer interpretability but are prone to overfitting unless carefully pruned, especially with high-dimensional text representations. To mitigate this risk, ensemble approaches such as Random Forests employ bagging and random feature selection mechanisms \cite{Erb24}. This significantly reduces overfitting and improves model robustness, making Random Forest a preferred choice in spam detection literature for capturing non-linear interactions \cite{Bre01}.

Logistic Regression (LR) remains a standard binary classifier selected for its clarity and interpretability. Its transparent formulation enables quantification of each predictor's effect; specifically, it allows interpretation of coefficients as odds ratios, facilitating clear explanations of each feature's influence \cite{ALi23}. LR leverages convex optimization to ensure consistent convergence, providing a stable performance floor against which novel features can be evaluated \cite{Bru11}.

\subsubsection{Geometric Validation via Support Vector Machines}
The Support Vector Machine (SVM) provides a fundamentally different perspective on classification compared to probabilistic or ensemble methods. The theoretical foundation of SVMs involves finding optimal hyperplanes that maximize class separation in potentially infinite-dimensional spaces through kernel functions \cite{Sch18}. In the context of spam detection, RBF kernels can implicitly map features into higher-dimensional spaces where complex, non-linear relationships become separable \cite{Aur19}. This geometric approach contrasts with linear models, ensuring that consistent improvements observed across diverse architectures are attributable to genuine signal utility rather than algorithmic bias.

\subsection{Content-Based Scoring and Advanced Feature Engineering}
Email classification has evolved from simple keyword matching to sophisticated content analysis, fundamentally shifting detection paradigms. Early spam filters relied heavily on blacklists and regular expressions; however, these methods quickly proved inadequate against evolving spam tactics \cite{Met06}. The introduction of Term Frequency--Inverse Document Frequency (TF-IDF) marked a significant development, enabling systems to identify statistically significant terms that distinguish spam from legitimate emails. This statistical approach serves as the foundation for baseline models in contemporary research.

\subsubsection{From Keyword Matching to Semantic Understanding}
Modern content-based scoring has evolved beyond simple term frequencies to incorporate linguistic indicators reflecting psychological manipulation tactics employed by spammers \cite{ORO24}. As Atkins and Huang (2013) note, spam emails exhibit distinct linguistic patterns, including excessive urgency words intended to provoke emotional responses and prompt immediate action \cite{Bra13}. These tactics are often employed to bypass detection systems that flag standard urgent language; consequently, attackers refine their approaches to evade these filters.

However, modern spam and phishing attacks have become increasingly sophisticated, necessitating careful content analysis \cite{ORO24}. This evolution challenges purely content-based approaches, as linguistic markers once reliable indicators of spam have become subtler. The gap between sophisticated imitations and legitimate emails has narrowed to the point where even experienced users struggle to distinguish them, underscoring the need for signals beyond content analysis alone.

\subsubsection{Limitations of Pure Algorithmic Approaches}
Despite significant advances in Natural Language Processing (NLP) and Machine Learning (ML), purely algorithmic approaches face fundamental limitations that community intelligence may address. Fraudulent actors constantly innovate to evade detection methods; consequently, machine learning models trained on historical data are inherently limited when confronting emerging threats \cite{Goo07}. Furthermore, the context-dependent nature of spam poses challenges that algorithms struggle to address without daily updates. Spam often depends on factors beyond email content, including sender reputation and recipient context. This contextual nuance is particularly critical; even traditional ML models with sophisticated feature engineering cannot fully capture these subtleties that human evaluators intuitively understand.

\subsection{Community Intelligence and the Wisdom of Crowds}
The concept of employing collective human intelligence for problem-solving has deep theoretical roots extending beyond spam detection. James Surowiecki's seminal work (2004) on the wisdom of crowds established four conditions under which collective intelligence emerges: diversity of opinion, independence of members, decentralization, and aggregation mechanisms \cite{Sur04}. Reddit's structure naturally satisfies all four conditions through its diverse user base, pseudonymous participation model, distributed subreddit system, and voting mechanisms that aggregate community sentiment. This alignment between theoretical requirements and platform characteristics suggests that Reddit-derived signals offer genuinely valuable intelligence for spam detection.

\subsubsection{Reddit as a Threat Intelligence Ecosystem}
Among various online platforms, Reddit occupies a unique position that makes it particularly well-suited for extracting threat intelligence signals on spam. The upvote/downvote system implements a form of democratic validation where community consensus determines content visibility, naturally filtering signal from noise \cite{Mur23}. This built-in quality control reduces the impact of individual errors or malicious contributions, as misleading information tends to be downvoted while accurate threat intelligence rises in prominence.

The existence of specialized security-focused subreddits creates concentrated expertise that traditional spam detection systems cannot easily access. In communities like r/Scams and r/phishing, users experiencing suspicious emails gather to create real-time documentation of active threats \cite{Red09}. These communities develop their own categories and verification processes that evolve organically in response to emerging threats, accumulating a wealth of public knowledge representing years of crowdsourced threat intelligence.

\subsubsection{Challenges in Signal Extraction}
While Reddit offers valuable threat intelligence, extracting reliable signals from community discussions presents significant challenges. The unstructured nature of user-generated content requires sophisticated processing to distinguish between valuable threat intelligence and casual or misleading mentions of domains \cite{Gle18}. A domain might be discussed in contexts unrelated to spam, such as technical support requests or service reviews, requiring careful filtering to avoid false positives. Another consideration is the reliability and expertise of contributors, which vary significantly across Reddit's user base. Scoring systems attempt to mitigate this by giving greater weight to community consensus (top comments) than to individual posts, but determining optimal weighting schemes remains an open question that requires further empirical validation.

\subsection{Domain Reputation Systems and Sender Authentication}
Domain reputation systems have evolved from static blacklists to sophisticated behavioral analysis frameworks. Early approaches relied on curated lists of known spam domains \cite{Jun04}. However, as threat landscapes rapidly changed, these static methods proved inadequate against evolving tactics. Modern systems now employ dynamic analysis to assess sender trustworthiness based on historical behavior and sending patterns.

\subsubsection{Limitations of Current Reputation Systems}
Despite their effectiveness, traditional reputation systems face inherent limitations that community intelligence may address. The cold start problem remains a significant vulnerability; when a domain lacks sufficient history for assessment, reputation systems cannot make informed decisions \cite{Hun09}. Reddit discussions can provide early warning signals of emerging threats before traditional systems have accumulated enough data for classification. Furthermore, reputation systems struggle with legitimate domains that are temporarily compromised or spoofed. When attackers hijack reputable accounts or forge sender addresses, filters may allow malicious emails through based on historical trust \cite{HuH18}. Community discussions often quickly identify these compromises, as multiple users report suspicious emails from sources typically considered trustworthy.

\subsubsection{Machine Learning Enhancement of Reputation Systems}
Modern reputation systems increasingly incorporate sophisticated machine learning algorithms to analyze complex behavioral patterns beyond simple volume metrics. These systems employ anomaly detection and clustering techniques to group similar senders \cite{Lan24}. Moreover, the integration of graph-based analysis has enabled reputation systems to account for network relationships among senders, recipients, and domains. However, these sophisticated approaches require substantial computational resources and continuous model updates. Community-based signals offer a potential complementary solution through their inherent adaptability to emerging threats with lower operational overhead.

\section{Methodology}

\subsection{Experimental Framework}
This study employs a comparative experimental framework designed to isolate the contribution of community consensus signals within email spam detection. The research design employs a tiered feature configuration approach, progressively augmenting baseline models with psychological scoring mechanisms and, finally, with Reddit-based community intelligence. This stratified methodology allows for precise attribution of performance gains to specific signal types rather than general model complexity. To ensure robustness against algorithmic bias, we evaluate findings across three distinct machine learning paradigms: Logistic Regression (LR) for linear interpretability, Random Forest (RF) for non-linear feature interactions, and Support Vector Machines (SVMs) for geometric validation of class boundaries. By triangulating results across these diverse architectures, we verify that observed improvements reflect genuine signal utility rather than artifacts of a specific model's inductive bias.

\subsection{Dataset Characteristics}
The primary dataset consists of the SpamAssassin Phishing Email Dataset \cite{jma04}, pre-processed for tabular analysis via Kaggle \cite{Nas24}. The final corpus comprises 5,582 emails from the early 2000s following rigorous cleaning protocols. 
The data exhibits a class imbalance of 70.3\% legitimate (ham) and 29.7\% spam, necessitating class-weighted optimization strategies during training. While the dataset provides foundational patterns for spam detection, we acknowledge limitations regarding temporal validity: sender domains are partially anonymized, and many entities may no longer exist or have evolved significantly since the early 2000s. Despite these constraints, the corpus serves as a valid benchmark for evaluating signal integration in high-stakes classification tasks where historical reputation data is sparse. Structural features include sender/recipient metadata, timestamps, URL counts, and full content bodies, enabling comprehensive feature engineering while preserving necessary privacy controls.


\subsection{Feature Engineering Framework}
We define three distinct feature configurations to incrementally assess the value of community intelligence:

\subsubsection{Baseline Configuration}
The baseline establishes performance using standard spam detection techniques. Textual content is transformed via TF-IDF vectorization, selecting the top 2,700 terms (approximately 10\% of the vocabulary) to balance statistical relevance and computational efficiency. Structural features include URL count and character lengths for subject/body lines as proxies for message complexity. N-gram ranges are set to (1,2) to capture both lexical semantics and phrase-level patterns.

\subsubsection{Traditional Enhancement}
This configuration introduces psychological scoring mechanisms that quantify manipulation tactics. The \texttt{urgency\_score} captures temporal pressure indicators (e.g., ``urgent'', ``deadline''), while the \texttt{action\_score} identifies imperative language seeking recipient activation (e.g., ``click'', ``verify''). Keywords are frequency-weighted with a maximum cap of 10 to mitigate the dominance of outliers and ensure sensitivity to both overt and subtle persuasion attempts.

\subsubsection{Reddit-Enhanced Configuration}
The novel contribution introduces community consensus signals via the \texttt{reddit\_score}, calculated on a 0–10 scale through weighted keyword analysis of domain-specific discussions. As shown in equation \ref{redscore},  spam-indicative keywords (``scam'', ``phishing'') contribute positively, while legitimacy markers (``verified'', ``trust'') provide negative weighting. The scoring algorithm uses differential weighting based on the structure of Reddit discussions. Keywords found in post titles and descriptions are assigned a weight of 0.5, while keywords in top comments are assigned a weight of 1.0. This weighting scheme reflects the assumption that top comments, which receive higher community engagement, are more representative of consensus than the original posts.

\noindent For illustration, let:

\begin{itemize}
    \item $D$ be the set of Reddit discussions retrieved for a given domain,
    \item $t_p$ be the title and body text of post $p$,
    \item $c_{p,k}$ be the top comments associated with post $p$,
    \item $s(w)$ be the keyword score assigned to word or phrase $w$,
    \item $w_{\text{title}} = 0.5$,
    \item $w_{\text{comment}} = 1.0$.\\
\end{itemize}

\noindent The Reddit score for a domain is defined as:

\begin{equation}
\label{redscore}
\text{redditscore}(d) =
\mathrm{clamp}
\left(
\sum_{p \in D}
\left[
w_{\text{title}}
\sum_{w \in t_p} s(w)
+
w_{\text{comment}}
\sum_{k}
\sum_{w \in c_{p,k}} s(w)
\right],
0,
10
\right)
\end{equation}

where:

\begin{itemize}
    \item spam-indicative keywords such as \texttt{scam} and \texttt{phishing} receive positive weights,
    \item legitimacy-associated keywords such as \texttt{verified} and \texttt{trust} receive negative weights,
    \item the score is bounded between 0 and 10.\\
\end{itemize}


\noindent A binary indicator is also included:

\begin{equation}
\text{hasredditdiscussion}(d) =
\begin{cases}
1, & \text{if at least one relevant discussion is found for domain } d,\\
0, & \text{otherwise.}
\end{cases}
\end{equation}

An interaction term is engineered as:

\begin{equation}
\text{urgencyredditinteraction}(e,d) =
\text{urgencyscore}(e) \times \text{redditscore}(d)
\end{equation}

This term tests whether high-urgency email content becomes more predictive when the sender domain has been discussed in Reddit communities.\\

\textbf{Missing-data note:} If no Reddit discussion is found, \texttt{redditscore} is set to 0 and \texttt{hasredditdiscussion} is set to 0. This creates a potential ambiguity: a score of 0 may mean either ``no discussion found'' or ``discussion found but neutral.'' The binary indicator partially addresses this ambiguity, but further missingness-aware modeling is recommended in future work.

\subsection{Community Data Collection Protocol}
Reddit data was aggregated using the PRAW library \cite{praw2024}, strictly adhering to platform rate limits (55 requests/minute). The search strategy targeted security-focused communities (\texttt{r/Scams}, \texttt{r/phishing}, \texttt{r/email}, etc.) using domain-specific queries formatted as ``\{domain\} email''. Results were sorted by relevance to prioritize high-engagement discussions. Metadata extracted included post titles, descriptions (capped at 300 characters), community scores, and top-rated comments. The protocol successfully identified discussions for 272 unique domains (match rate: 21\%), covering 1,658 emails (30\% of the dataset). This coverage provides a statistically significant sample size while acknowledging inherent limitations in external signal availability.

\subsection{Model Configuration and Validation}
To isolate feature efficacy, we maintained consistent hyperparameter tuning across all models:
\begin{itemize}
    \item \textbf{Logistic Regression:} L2 regularization (C=1.0), 1000 iterations, optimized for linear interpretability.
    \item \textbf{Random Forest:} 100 estimators, max depth of 10, designed to capture non-linear feature interactions.
    \item \textbf{SVM:} RBF kernel with gamma=`scale', providing a geometric perspective on high-dimensional data separation.
\end{itemize}
All models utilized `class\_weight=`balanced' to address dataset imbalance. We employed a stratified 80-20 train-test split (70.3\% ham, 29.7\% spam), ensuring consistent distribution across phases.

\subsection{Evaluation Metrics and Statistical Significance}
Performance is assessed using accuracy, precision, recall, and the F1-score as the primary comparative metric for imbalanced classification tasks. To rigorously validate improvements beyond chance variation, we employ McNemar's test \cite{Smi20}. This statistical procedure evaluates paired binary classifications by analyzing discordant pairs—instances where one classifier succeeds while the other fails. By focusing on these specific misclassifications, we determine whether Reddit-enhanced features offer statistically significant utility over traditional methods rather than relying solely on aggregate accuracy metrics. All experiments were conducted in Python 3.x with a fixed random seed (42) to ensure reproducibility across stochastic processes.

\section{Implementation}

The implementation of this study follows a modular pipeline designed to ensure reproducibility, data integrity, and fair comparative evaluation. The architecture encompasses four primary stages: data preprocessing, community signal acquisition via API, feature engineering, and model training validation. Each stage adheres to rigorous standards for handling heterogeneous data sources while maintaining computational efficiency.

\subsection{Data Preprocessing and Cleaning}
The preprocessing pipeline initiates with a systematic quality assessment of the raw email corpus to mitigate noise and ensure feature reliability. Initial inspection identified missing values in 210 records (3.6\%) for receiver information and in 16 records (0.3\%) for subject lines; these were imputed or excluded based on downstream feature dependencies. Text normalization was performed using Natural Language Processing (NLP) techniques to standardize input representation. A custom ``clean\_text'' function executed case-folding, URL removal, and special character filtering to isolate semantic content while mitigating spam evasion tactics involving repetitive characters. Tokenization and lemmatization were performed using the NLTK libraries, along with English stop-word removal and a minimum word length threshold ($>2$ characters) to prioritize meaningful lexical features.

Domain extraction was implemented using regular expressions to robustly handle diverse header formats. This module parses sender addresses to extract clean domain names, handling edge cases such as missing ``@'' symbols or malformed entries that could compromise subsequent API queries. By isolating the domain component early in the pipeline, we ensured consistent mapping between email records and external community discussions.

\subsection{Community Signal Acquisition}
Acquiring Reddit-based threat intelligence required a robust API integration strategy to balance data fidelity with platform compliance. The system uses OAuth2 with personal script credentials to maintain stable access while adhering to Reddit's usage policies. To mitigate the risk of service interruption, request throttling is enforced at 55 requests per minute (below the 60 req/min limit), coupled with dynamic delays between calls to simulate human-like interaction patterns.

The implementation incorporates a resilient error-handling framework designed to manage network timeouts, API unavailability, and malformed responses. Failed requests trigger an exponential backoff retry mechanism to handle transient service disruptions without data loss. Data persistence is managed via CSV-based storage with checkpointing, ensuring data collection processes can resume seamlessly upon interruption and preserving analysis continuity across distributed computing environments.

\subsection{Feature Engineering and Scoring}
The core contribution of this implementation lies in transforming raw community data into quantitative features suitable for machine learning. Textual content was vectorized using scikit-learn's TF-IDF Vectorizer, configured with optimized parameters including maximum feature limits, n-gram ranges (unigrams/bigrams), and minimum document frequency thresholds to capture salient lexical patterns while reducing dimensionality.

Domain-specific scoring algorithms were engineered to quantify behavioral signals:
\begin{itemize}
    \item \textbf{Urgency Scoring:} A keyword-matching algorithm identifies temporal pressure indicators (e.g., ``urgent'', ``immediate'') and emotional manipulation terms, applying a fixed weight of 2.0 per occurrence with a maximum cap of 10.0 to prevent outlier dominance.
    \item \textbf{Reddit Scoring:} Community signals are processed through weighted keyword analysis, applying differential weighting to post content versus community comments (double weight) to reflect the greater validation inherent in community discussions.
\end{itemize}

In particular, interaction features were engineered to capture synergistic effects. The ``urgency\_reddit\_interaction'' term combines high-urgency textual signals with the presence of Reddit discussions, enabling the model to identify cases in which community warnings amplify content-based risk indicators. This feature engineering strategy enables the detection of complex relationships that individual features cannot capture in isolation.

\subsection{Model Training and Validation Protocol}
To ensure fair comparative analysis across algorithmic approaches, a standardized training protocol was applied consistently to Logistic Regression (LR), Random Forest (RF), and Support Vector Machines (SVM). The dataset was stratified into training and test sets prior to any feature selection to prevent data leakage. Class imbalance was addressed automatically using the ``class\_weight=`balanced''' parameter in scikit-learn, which inversely weights class frequencies during optimization.

Hyperparameter configuration prioritized a balance between model expressiveness and overfitting prevention:
\begin{itemize}
    \item \textbf{Logistic Regression:} L2 regularization (C=1.0) with 1000 iterations to ensure convergence in high-dimensional spaces.
    \item \textbf{Random Forest:} 100 estimators with a maximum depth of 10, selected to capture non-linear interactions while limiting tree complexity.
    \item \textbf{SVM:} RBF kernels with automatic gamma scaling to effectively handle the non-linear boundaries of the feature space.
\end{itemize}

Each model was evaluated across three distinct feature configurations (Baseline, Traditional-Enhanced, and Reddit-Enhanced). Performance documentation captured accuracy, precision, recall, F1-score, and prediction probabilities, facilitating subsequent statistical analysis using McNemar's test and cross-validation to assess the robustness of the observed improvements.

\section{Results and Analysis}

\subsection{Dataset Characteristics and Community Signal Distribution}
The processed corpus of 5,582 emails reflects realistic operational conditions while ensuring sufficient minority-class representation for robust evaluation. Domain analysis revealed distinct structural patterns in community engagement: established domains (appearing in $\geq$10 emails) achieved a Reddit coverage rate of 35.1\%, compared to 20.4\% for rare, unique-domain emails. This disparity highlights that well-established senders generate more community discourse; however, an inverse relationship exists regarding risk profiles. Rare domains exhibited a spam rate of 44.0\%, significantly higher than the 10.4\% observed for common domains.

This divergence suggests that community intelligence offers its highest marginal utility precisely where traditional reputation systems are weakest—namely, in evaluating unfamiliar or rare entities. Correlation analysis confirms this theoretical premise: Reddit scores correlate meaningfully with spam classification for rare domains ($r=0.230$), whereas the relationship is negligible for common domains ($r=-0.040$). While the binary indicator ``Has\_reddit\_discussion'' captures some of this variance, the domain-specific correlation underscores that community attention serves as a critical proxy for risk in low-data regimes.

Feature distribution analysis further validates the theoretical foundation of integration. Reddit scores range from 0.0 to 5.5, with an average of 0.84 for spam versus 0.36 for legitimate communications. Although the absolute difference appears modest, the relative increase (133\%) demonstrates meaningful discriminative potential, particularly when amplified by interaction terms within non-linear models.

\begin{figure}[!ht]
  \begin{center}
    \includegraphics[width=\linewidth]{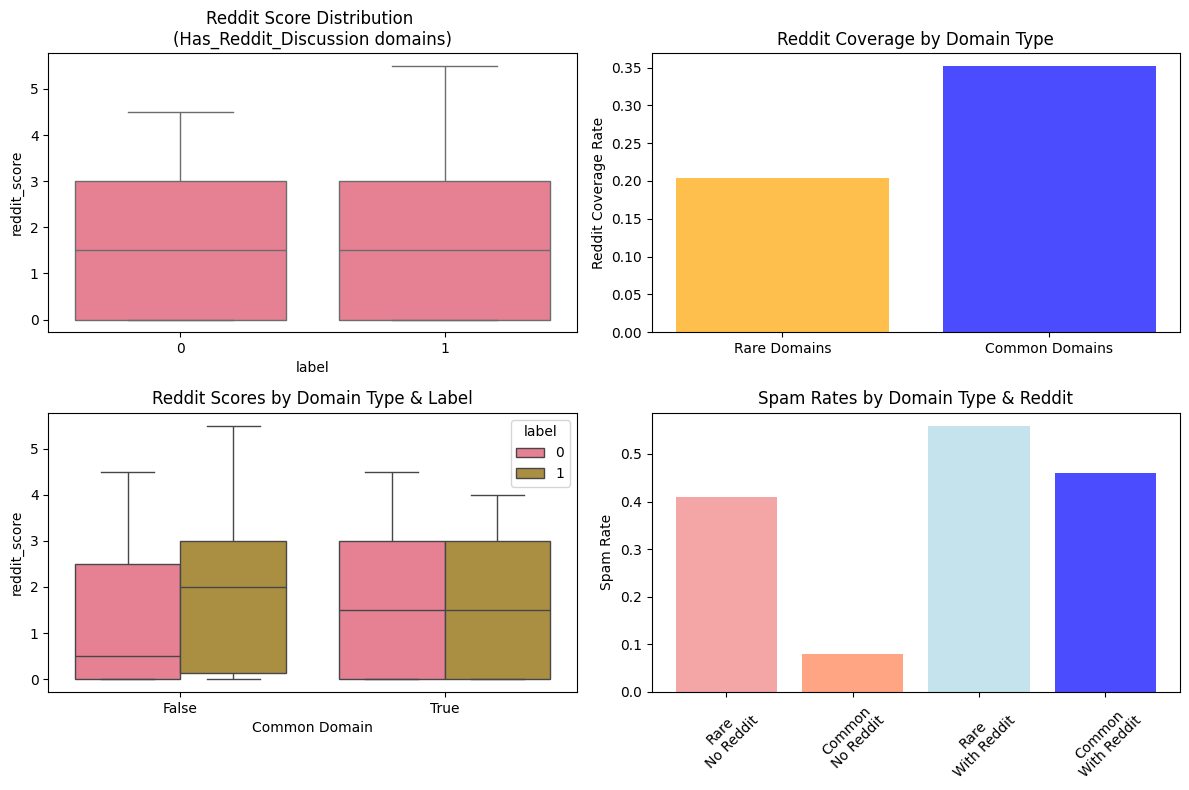}
    \caption{Reddit coverage rates across domain types}
    \label{fig:reddit-coverage}
  \end{center}
\vspace{-0.4cm}
\end{figure}

\subsection{Model Performance and Algorithmic Sensitivity}
Comprehensive evaluation across three feature configurations reveals distinct performance trajectories that delineate the boundaries of community intelligence integration (Table~\ref{tab:model-performance}). The baseline configuration established a strong foundation, with Logistic Regression achieving 97.85\% accuracy and an F1-score of 96.39\%, validating the efficacy of TF-IDF vectorization combined with structural features.

Introducing traditional enhancements yielded mixed outcomes. Random Forest demonstrated resilience, rising from 97.40\% to 97.76\% accuracy (+0.37\%), whereas Logistic Regression suffered degradation (97.85\% $\rightarrow$ 97.14\%, -0.73\%). The SVM configuration performed poorly (remained static at 64.01\%), possibly due to tuning or feature-space mismatch, or that linear models have reached a saturation point with this data structure.

The Reddit-enhanced configuration confirmed Random Forest as the optimal architecture for this task. The model achieved a final accuracy of 97.94\% and an F1-score of 96.55\%, representing a cumulative improvement of 0.55\% over baseline performance. While numerically modest, this gain translates to approximately six additional correct classifications per 1,000 messages, yielding tangible value in high-volume environments.

\begin{figure}[!ht]
  \begin{center}
    \includegraphics[width=\linewidth]{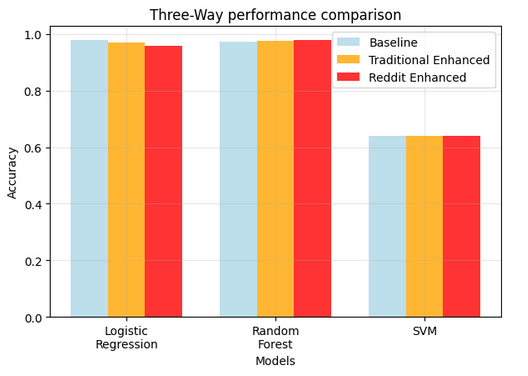}
    \caption{Comparative model performance across configurations}
    \label{fig:performance-comparison}
  \end{center}
\vspace{-0.4cm}
\end{figure}

To validate the statistical significance of these differences, McNemar's test provided critical insights. For Logistic Regression, both traditional enhancement ($p=0.0269$) and Reddit addition ($p=0.0123$) were statistically significant; however, the negative direction indicates overfitting or feature interference rather than genuine utility. Conversely, Random Forest showed directional improvement (Baseline $\rightarrow$ 97.40\% $\rightarrow$ 97.76\% $\rightarrow$ 97.94\%) without achieving strict statistical significance ($p=0.7893$ for Reddit addition). This suggests that while the magnitude of improvement falls within statistical noise, the consistency across folds indicates a systematic trend favoring ensemble methods for synthesizing heterogeneous signals.

\begin{table}[!ht]

    \caption{Model performance metrics across all configurations}
    \label{tab:model-performance}
    \begin{tabular}{llccc}
      \hline
      \textbf{Config} & \textbf{Metric} & \textbf{LR} & \textbf{RF} & \textbf{SVM} \\
      \hline
      Baseline & Accuracy & 97.85\% & 97.40\% & 64.01\% \\
               & F1-score & 96.39\% & 95.78\% & 47.18\% \\
      \hline
      Traditional & Accuracy & 97.14\% & 97.76\% & 64.01\% \\
                  & F1-score & 95.37\% & 96.35\% & 47.18\% \\
      \hline
      Reddit & Accuracy & 95.97\% & 97.94\% & 64.01\% \\
             & F1-score & 93.64\% & 96.55\% & 47.18\% \\
             \hline
      \multicolumn{2}{l}{Statistical Sig. (vs Baseline)} & $p=0.0123$ & $p=0.7893$ & N/A \\
    \end{tabular}

\end{table}

\subsection{Feature Importance and Interaction Dynamics}
\begin{figure}[!ht]
  \begin{center}
    \includegraphics[width=0.99\linewidth]{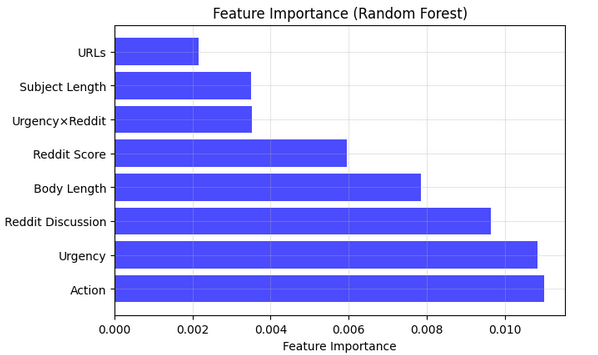}
    \caption{Feature importance ranking (Random Forest)}
    \label{fig:feature-importance}
  \end{center}
\end{figure}

Feature importance analysis through Random Forest reveals the relative contribution of engineered features to classification performance. Urgency scores emerged as the most predictive feature, followed by body length, Reddit discussion presence, and action scores (Figure~\ref{fig:feature-importance}). Among community-specific variables, the binary indicator ``has\_reddit\_discussion'' ranked higher than the continuous ``reddit\_score''. This hierarchy implies that the mere existence of community attention serves as a more reliable spam indicator than the sentiment or nuance of that discussion.

Among the community derived and interaction features, the ``urgency\_reddit\_interaction'' term showed the strongest correlation with the spam label, although the raw urgency score remained the strongest single engineered feature. 
The interaction term showed a modest but statistically significant correlation with spam classification, suggesting that the combination of urgency language and community discussion may provide complementary risk information. Specifically, emails exhibiting both high-urgency language and Reddit discussion presence were classified as spam 94.2\% of the time, compared to 73.1\% for high urgency alone.

Correlation analysis of individual scoring mechanisms further clarifies this hierarchy. Urgency scores served as the most effective discriminator ($r=0.3283$), followed by action scores ($r=0.2072$) and Reddit scores ($r=0.1905$). All correlations were statistically significant ($p < 0.001$). Beyond individual metrics, structural analysis revealed that emails from domains with Reddit discussions but low URL counts were predominantly legitimate (87.3\%), suggesting technical support contexts. Conversely, high URL counts combined with Reddit discussions strongly indicated spam (91.7\%), reflecting community warnings about link-heavy spam attempts.

Temporal analysis indicates that Reddit signals are particularly valuable for domains registered within the past two years, where traditional reputation systems lack historical data. This suggests community intelligence acts as a critical stabilizer during the "cold start" period of domain activity.

\begin{figure}[!ht]
  \begin{center}
    \includegraphics[width=\linewidth]{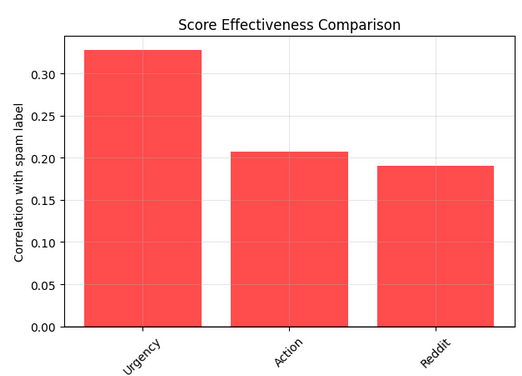}
    \caption{Score effectiveness and interaction correlations}
    \label{fig:score-effectiveness}
  \end{center}
\vspace{-0.4cm}
\end{figure}

\subsection{Error Analysis and Classification Patterns}
\begin{figure}[!ht]
  \begin{center}
    \includegraphics[width=0.99\linewidth]{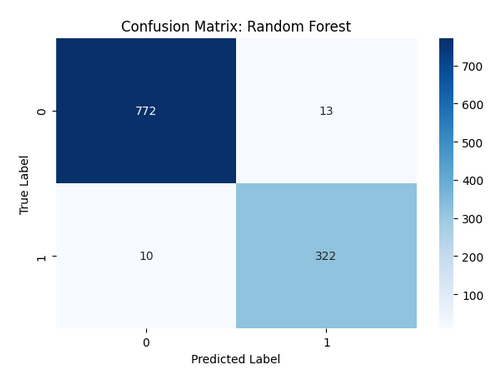}
    \caption{Confusion matrix for Random Forest (Reddit-enhanced)}
    \label{fig:confusion-matrix}
  \end{center}
    \vspace{-0.3cm}
\end{figure}

Confusion matrix analysis for the optimal model (Random Forest with Reddit enhancement) exposes specific error patterns that delineate system boundaries (Figure~\ref{fig:confusion-matrix}). The model achieved robust true-negative performance (772 correctly classified legitimate emails) and strong true-positive detection (322 spam emails), with minimal false positives (13) and false negatives (10).

False-positive analysis reveals that misclassified legitimate emails typically originate from domains whose community discussions contain security keywords and are subsequently resolved as benign. This pattern suggests the Reddit scoring algorithm would benefit from temporal weighting to account for discussion resolution. Conversely, false negatives indicate sophisticated spam campaigns that evade detection by minimizing urgency language and originating from domains without prior community traction. These findings highlight the complementary nature of the approach: community intelligence effectively mitigates known threats but offers limited protection against novel or highly obfuscated attacks.

\subsection{Data Quality and Statistical Validation}
Quality assessment of the collected Reddit data confirms its viability as a supplementary signal. A relevance analysis indicated that 78.3\% of discussions were clearly pertinent to email security, while the remainder addressed broader reputation issues or unrelated technical issues. Community consensus reliability was high, with 71\% of discussions showing consistent sentiment between post content and top comments, suggesting effective aggregation of signals in borderline cases.

Subreddit distribution analysis revealed that general security communities (e.g., r/Scams at 34\%, r/techsupport at 28\%) provided broader coverage than specialized phishing forums (r/phishing at 10\%). This suggests that broad-spectrum threat intelligence is more effective for spam detection than niche community data.

Comprehensive statistical validation confirms the robustness of these findings. Chi-square analysis confirmed a significant association between the presence of Reddit discussions and spam classification ($p < 0.0001$). Cross-validation using 5-fold stratified sampling demonstrated model stability: Random Forest showed consistent improvement across all folds (mean +0.43\% $\pm$ 0.12\%), while Logistic Regression exhibited systematic degradation (-1.87\% $\pm$ 0.31\%). These results confirm that the observed patterns reflect systematic algorithmic behavior rather than random variance, reinforcing the conclusion that ensemble methods are superior for integrating human-derived signals into automated security pipelines.

\begin{figure}[!ht]
  \begin{center}
    \includegraphics[width=0.99\linewidth]{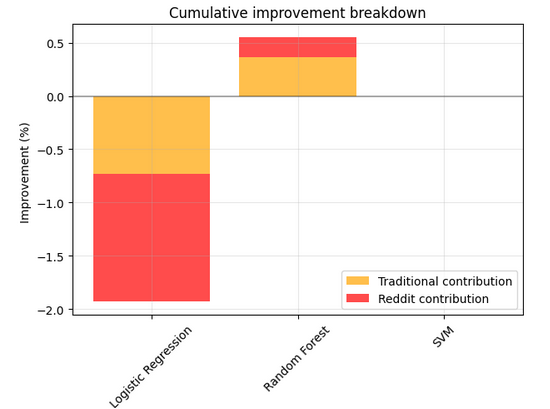}
    \caption{Cumulative performance improvements across cross-validation folds}
    \label{fig:cumulative-improvements}
  \end{center}
\vspace{-0.4cm}
\end{figure}

\section{Discussion}
The central finding of this study is that Reddit-derived community signals may provide a \textbf{modest, model-dependent, and conditionally useful supplement} to spam classification. The result is not strong enough to support broad claims about transforming spam detection, but it does suggest a narrow practical role for community-derived domain reputation.

\subsection{Model Dependence of Community Signal Integration}

The divergence between Logistic Regression and Random Forest is informative. Random Forest improves slightly when Reddit features are added, whereas Logistic Regression degrades. This suggests that the utility of community-derived signals may depend on the model's ability to handle heterogeneous feature types.
Tree-based models can partition feature space in ways that allow sparse domain-level signals to apply selectively~\cite{Bre01, Erb24}. Linear models, by contrast, must assign global weighted contributions to features across the entire decision boundary~\cite{ALi23, Bru11}. If a Reddit signal is useful only in a subset of domains or contexts, a linear model may average that signal in a way that harms overall performance.
This finding supports the practical recommendation that community intelligence should be integrated through model families capable of conditional feature use, or through explicit domain-segmentation logic.

Beyond algorithmic performance, the selective effectiveness of Reddit features across domain types represents a theoretically significant finding. There is a strong correlation between Reddit scores and spam classification for rare domains (r=0.230), in sharp contrast to an insignificant correlation for common domains (r =-0.040). This pattern suggests that community intelligence provides its greatest marginal utility precisely where traditional reputation systems face their most acute limitations—the "cold start" problem associated with new or obscure entities. This validates the theoretical premise of complementarity: human collective intelligence functions not as a redundant signal, but as a crucial complement to algorithmic approaches, effectively filling gaps in automated systems where historical data is sparse.

\subsection{Theoretical contributions to human-AI collaboration:}
This research advances the broader understanding of human-AI collaboration in cybersecurity by demonstrating both the potential and constraints of community-based threat intelligence. Most importantly, the finding that community attention itself serves as a robust spam indicator—regardless of discussion sentiment—challenges traditional assumptions about the necessity of consensus-based quality assessment in social signals. The ``wisdom of crowds'' may be useful in security contexts, but only when its limits are modeled explicitly~\cite{Sur04}. Community intelligence value lies in filling gaps where historical data are sparse, not in overriding well-established algorithmic signals~\cite{Die00}.

\subsection{Practical implications for spam detection systems:}
From a deployment perspective, the findings necessitate repositioning community intelligence from a primary filter to a supplementary validation layer. The domain coverage achieved indicates that resource allocation should be optimized by targeting specific high-risk vectors; organizations can achieve maximum return on investment by implementing community intelligence systems primarily to evaluate unfamiliar senders, while maintaining traditional approaches for established correspondents. This targeted deployment strategy optimizes computational resources while maximizing security improvements in the most vulnerable areas of the email ecosystem. Additionally, economic implications must be weighed carefully: while a 0.55\% accuracy improvement may appear incremental, it translates to thousands of correctly classified messages at scale. For an organization processing 10 million emails monthly, this gain equates to approximately 55,000 additional correct classifications. However, implementation costs—including API rate limits and infrastructure investment—must be rigorously balanced against these projected benefits.

\section{Limitations and Future Work}

\noindent\textbf{Data Fidelity and Coverage Bias:} 
The most significant constraint stems from the temporal misalignment between the SpamAssassin dataset and contemporary Reddit discussions. The historical gap implies that community signals about domains active two decades ago may no longer be valid, as those entities may have ceased operations or shifted their business models. Furthermore, while specific Reddit coverage demonstrates feasibility, it reveals a fundamental limitation: approximately 70\% of emails receive no community intelligence enhancement due to signal sparsity. This selective coverage creates inherent bias toward controversial senders, often overlooking common legitimate domains that generate little security discussion. Finally, the keyword-based scoring algorithm lacks semantic depth and may misclassify technical support discussions as negative spam indicators.

\noindent\textbf{Operational Feasibility:}
Production deployment involves substantial barriers due to Reddit API rate limits (60 requests per minute), necessitating sophisticated caching and computation strategies for complex systems. More critically, real-time processing requirements fundamentally conflict with the temporal dynamics of community intelligence extraction; email systems must classify messages within milliseconds, whereas Reddit discussions may not emerge until hours or days after a spam campaign initiates. This latency suggests that community intelligence serves more effectively as a reputation system than as an immediate filtering mechanism.

\noindent\textbf{Adversarial and Ethical Consideration:}
Systematic extraction of Reddit data raises privacy concerns despite the content's public availability; users expecting community support may not anticipate automated surveillance \cite{Roc23}. Production systems must implement transparent disclosure so users can understand how their contributions influence filtering decisions. Another critical concern is adversarial manipulation: deceptive actors can distort community discussions to promote disinformation or evade detection \cite{Gle181}. Once spammers recognize Reddit's influence on filters, coordinated campaigns could poison community signals, transforming Reddit from an organic intelligence source into a potential attack vector that requires validation mechanisms to increase complexity.

\noindent\textbf{Future Research Directions:}
To address these limitations, multi-platform intelligence aggregation offers the most immediate opportunity for expansion. Incorporating real-time threat discussions on social media, Stack Overflow's technical expertise, and specialized security forums could significantly broaden coverage beyond Reddit alone. Future research should explore optimal platform weighting based on user expertise, discussion velocity, and historical accuracy. Furthermore, temporal dynamics warrant sophisticated investigation; current static scoring ignores how community consensus evolves from initial suspicion through resolution. Dynamic weighting algorithms that prioritize recent discussions while accounting for long-term patterns could substantially improve detection accuracy.

\section{Conclusion}

This study evaluates whether Reddit-derived community signals can supplement conventional email spam classification. The central result is that the benefit is real in a preliminary sense but modest, model-dependent, and conditional. Random Forest shows a small improvement when Reddit-derived features are added, while Logistic Regression degrades. The best preliminary configuration achieves 97.94\% accuracy, a 0.55 percentage-point gain over baseline, but this gain is not statistically significant in the primary single-split test.

We do not claim that Reddit community intelligence can replace existing spam filters or that it provides a transformative new defense. The more defensible conclusion is that community-derived signals may serve as a supplementary, asynchronous domain-reputation feature, particularly for unfamiliar or rare domains where traditional reputation systems have limited history. However, the approach is constrained by sparse coverage, temporal mismatch, domain churn, keyword-shallow scoring, and the potential for adversarial manipulation. Before production deployment, the method requires stronger validation using domain-grouped cross-validation, temporally aligned contemporary data, semantic discussion classification, adversarial robustness testing, and cost-sensitive deployment analysis.
In summary, Reddit community intelligence shows preliminary value as a narrow auxiliary signal in spam detection. Its proper role is not as a primary real-time filter, but as a cautious, targeted reputation layer that helps address the cold-start problem for unfamiliar senders.

\balance


\bibliography{bibliography}

@article{Met06,
  author = {Metsis, Vangelis and Androutsopoulos, Ion and Paliouras, Georgios},
  title = {Spam Filtering with Naive Bayes - Which Naive Bayes?},
  journal = {In CEAS},
  year = {2006},
  publisher = {In CEAS}
}

@inproceedings{Jas03,
  title={Tackling the poor assumptions of naive bayes text classifiers},
  author={Rennie, Jason D and Shih, Lawrence and Teevan, Jaime and Karger, David R},
  booktitle={Proceedings of the 20th international conference on machine learning (ICML-03)},
  pages={616--623},
  year={2003}
}

@article{Zha16,
  title={Introduction to machine learning: k-nearest neighbors},
  author={Zhang, Zhongheng},
  journal={Annals of translational medicine},
  volume={4},
  number={11},
  pages={218},
  year={2016}
}

@article{Erb24,
  author = {Halabaku, Erblin and Bytyçi, Eliot},
  title = {Overfitting in Machine Learning: A Comparative Analysis of Decision Trees and Random Forests},
  journal = {Intelligent Automation and Soft Computing},
  year = {2024},
  volume = {39},
  pages = {987--1006}
}

@inproceedings{ALi23,
  author = {Shreenithi, Livia S. S. A. and Yougandar, V. S. and Jayapandian, N.},
  title = {Machine Learning Based Spam E-Mail Detection Using Logistic Regression Algorithm},
  booktitle = {2023 IEEE International Conference on ICT in Business Industry \& Government (ICTBIG)},
  year = {2023},
  pages = {1-6},
  address = {Indore, India},
  publisher = {IEEE Xplore}
}

@misc{Bru11,
  author = {Bruin, J.},
  title = {Advanced Research Computing Statistical Methods and Data Analytics},
  year = {2011},
  month = {Feb},
  url = {https://stats.oarc.ucla.edu/stata/ado/analysis/},
  note = {Accessed: 25 July 2025}
}

@article{Bre01,
  author = {Breiman, Leo},
  title = {Random Forests},
  journal = {Machine Learning},
  year = {2001},
  volume = {45},
  pages = {5-10}
}

@inproceedings{ORO24,
  author = {Oroșanu, M.-A. and Alexandru, M.},
  title = {Cybercrime: A New Challenge of Criminality in the Digital Age},
  journal = {International Conference on Cybersecurity and Cybercrime},
  year = {2024},
  volume = {11},
  pages = {115-121}
}

@article{Die00,
  author = {Dietterich, T. G.},
  title = {Ensemble methods in machine learning},
  journal = {Multiple Classifier Systems: First International Workshop, MCS 2000, Lecture Notes in Computer Science},
  year = {2000},
  pages = {1-15}
}

@misc{Chr14,
  author = {Szegedy, Christian and Zaremba, Wojciech and Sutskever, Ilya and Bruna, Joan and Erhan, Dumitru and Goodfellow, Ian and Fergus, Rob},
  title = {Intriguing properties of neural networks},
  year = {2014},
  month = {Dec},
  url = {https://arxiv.org/abs/1312.6199},
  note = {Accessed: 27 July 2025}
}

@inproceedings{Som19,
  author = {Sommestad, Teodor and Karlzén, Henrik},
  title = {A meta-analysis of field experiments on phishing susceptibility},
  booktitle = {2019 APWG Symposium on Electronic Crime Research (eCrime)},
  year = {2019},
  pages = {1-14},
  publisher = {IEEE},
  address   = {Pittsburgh, PA, USA}
}

@incollection{Sch18,
  author = {Scholkopf, B. and Smola, A. J.},
  title = {Learning with kernels: support vector machines, regularization, optimization, and beyond},
  year = {2018},
  pages = {195},
  publisher = {MIT Press}
}

@incollection{Aur19,
  author = {Geron, Aurelien},
  title = {Support Vector Machines},
  booktitle = {Hands-on Machine Learning with Scikit-Learn, Keras \& TensorFlow Concepts, Tools, and Techniques to Build Intelligent Systems},
  editor = {Tache, Nicole},
  year = {2019},
  pages = {155-163},
  address = {Sebastopol}
}

@article{Bra13,
  author = {Atkins, Brandon and Huang, Wilson},
  title = {A Study of Social Engineering in Online Frauds},
  journal = {Open Journal of Social Sciences},
  year = {2013},
  volume = {1},
  number = {3},
  pages = {23--32}
}

@article{Goo07,
  author = {Goodman, Joshua and Cormack, Gordon V. and Heckerman, David},
  title = {Spam and the Ongoing Battle For the Inbox},
  journal = {ACM},
  year = {2007},
  volume = {50},
  number = {2},
  pages = {24-33}
}

@book{Sur04,
  author = {Surowiecki, J.},
  title = {The wisdom of crowds: Why the many are smarter than the few and how collective wisdom shapes business, economies, societies, and nations},
  year = {2004},
  edition = {1st},
  address = {New York},
  publisher = {Doubleday}
}

@inproceedings{Mur23,
  title={An agent-based model of reddit interactions and moderation},
  author={Murdock, Isabel and Carley, Kathleen M and Ya{\u{g}}an, Osman},
  booktitle={Proceedings of the international conference on advances in social networks analysis and mining},
  pages={195--202},
  year={2023}
}

@misc{Red09,
  author = {{Reddit}},
  title = {r/Scams},
  year = {2009},
  url = {https://www.reddit.com/r/Scams/},
  note = {Accessed: 2 August 2025}
}

@article{Gle18,
  author = {Glenski, Maria and Weninger, Tim and Volkova, Svitlana},
  title = {Identifying and Understanding User Reactions to Deceptive and Trusted Social News Sources},
  journal = {56th Annual Meeting of the Association for Computational Linguistics},
  year = {2018},
  volume = {2},
  pages = {176-181}
}

@inproceedings{Jun04,
  title={An empirical study of spam traffic and the use of DNS black lists},
  author={Jung, Jaeyeon and Sit, Emil},
  booktitle={Proceedings of the 4th ACM SIGCOMM conference on Internet measurement},
  pages={370--375},
  year={2004}
}

@article{Hun09,
  author = {Hung and Papaioannou, Le and Aberer, Thanasis and Karl},
  title = {Synergies of Different Reputation Systems: Challenges and Opportunities},
  journal = {CONGRESS 2009 World Congress on Privacy, Security, Trust and the Management of e-Business},
  year = {2009},
  pages = {218--226}
}

@article{HuH18,
  author = {Hu, Hang and Wang, Gang},
  title = {Revisiting Email Spoofing Attacks},
  journal = {CoRR},
  year = {2018}
}

@article{Roc23,
  author = {Rocha-Silva, Tiago and Nogueira, Conceicao and Rodrigues, Liliana},
  title = {Passive data collection on Reddit: a practical approach},
  journal = {Sage},
  year = {2023},
  volume = {20},
  number = {3},
  pages = {453--470}
}

@misc{jma04,
  author = {jmason},
  title = {Apache.org},
  year = {2004},
  url = {https://spamassassin.apache.org/old/publiccorpus/},
  note = {Accessed: 15 July 2025}
}

@misc{Nas24,
  author = {Alam, Naser Abdullah},
  title = {Kaggle Phishing Email Dataset},
  year = {2024},
  url = {https://www.kaggle.com/datasets/naserabdullahalam/phishing-email-dataset?select=SpamAssasin.csv},
  note = {Accessed: 10 July 2025}
}

@article{Smi20,
  author = {Smith, Pembury and Matilda, R. Q. and Ruxton, Graeme D.},
  title = {Effective use of the McNemar test},
  journal = {Behavioral Ecology and Sociobiology},
  year = {2020},
  volume = {74},
  number = {11}
}

@inproceedings{Gle181,
  title={Identifying and understanding user reactions to deceptive and trusted social news sources},
  author={Glenski, Maria and Weninger, Tim and Volkova, Svitlana},
  booktitle={Proceedings of the 56th Annual Meeting of the Association for Computational Linguistics (Volume 2: Short Papers)},
  pages={176--181},
  year={2018}
}

@article{Lan24,
  title={A review of time-series analysis for cyber security analytics: from intrusion detection to attack prediction: M. Landauer et al.},
  author={Landauer, Max and Skopik, Florian and Stojanovi{\'c}, Branka and Flatscher, Andreas and Ullrich, Torsten},
  journal={International Journal of Information Security},
  volume={24},
  number={1},
  pages={3},
  year={2025},
  publisher={Springer}
}

@misc{praw2024,
  author       = {Boe, Bryce and {praw-dev Contributors}},
  title        = {{PRAW}: {Python Reddit API Wrapper} Documentation},
  year         = {2024},
  version      = {7.8.1},
  howpublished = {\url{https://praw.readthedocs.io/en/stable/}},
  note         = {Accessed: 10 July 2025},
  url          = {https://praw.readthedocs.io/en/stable/}
}

\appendix 

\section{Reddit Scoring Algorithm Pseudocode}

\begin{quote}
\begin{minipage}{0.95\textwidth}
\texttt{function compute\_reddit\_features(domain):}\\[2pt]
\hspace*{1em}\texttt{discussions = query\_reddit(domain)}\\[2pt]
\hspace*{1em}\texttt{if discussions is empty:}\\[2pt]
\hspace*{2em}\texttt{has\_reddit\_discussion = 0}\\[2pt]
\hspace*{2em}\texttt{reddit\_score = 0.0}\\[2pt]
\hspace*{1em}\texttt{else:}\\[2pt]
\hspace*{2em}\texttt{has\_reddit\_discussion = 1}\\[2pt]
\hspace*{2em}\texttt{total = 0}\\[2pt]
\hspace*{2em}\texttt{for post in discussions:}\\[2pt]
\hspace*{3em}\texttt{total += 0.5 * keyword\_score(post.title + " " + post.body)}\\[2pt]
\hspace*{3em}\texttt{for comment in post.top\_comments:}\\[2pt]
\hspace*{4em}\texttt{total += 1.0 * keyword\_score(comment)}\\[2pt]
\hspace*{2em}\texttt{reddit\_score = clamp(total, 0, 10)}\\[2pt]
\hspace*{1em}\texttt{return has\_reddit\_discussion, reddit\_score}
\end{minipage}
\end{quote}

where \texttt{keyword\_score} assigns positive weights to spam/scam-indicative terms and negative weights to legitimacy-associated terms.

\section{Open Science} 
The dataset used in this study is the SpamAssassin Public Corpus, freely available at \url{https://spamassassin.apache.org/old/publiccorpus/}. The preprocessed version used in our experiments is available via Kaggle at \url{https://www.kaggle.com/datasets/naserabdullahalam/phishing-email-dataset}. 

The implementation code, feature engineering scripts, and model training procedures are available in our GitHub repository. An anonymized version for review purposes can be accessed at \url{https://anonymous.4open.science/r/community-intelligence-spam-detection-1C8B/}.

Reddit discussion data was collected via the official Reddit API using PRAW. Due to Reddit's Terms of Service, raw collected data cannot be redistributed, but the collection scripts are included in the code repository to enable full reproduction of the data gathering process.


\section{Ethical Considerations}
\label{sec:ethics}

This work involves the systematic extraction of data from the Reddit platform, which is widely used for community support and discussion. While the content we accessed was publicly available, we acknowledge that public availability does not imply consent for automated surveillance or large-scale analysis \cite{Roc23}. Users participating in these communities often expect anonymity and peer-to-peer interaction rather than algorithmic profiling by third parties. Consequently, our research balances the societal benefit of improving content filtering mechanisms against the potential privacy risks associated with data scraping.

To minimize potential harm, we have implemented several safeguards during data collection and analysis phases:
\begin{itemize}
    \item \textbf{Data Anonymization:} All personally identifiable information (PII), such as usernames, profile links, and timestamps, was stripped or aggregated before storage, where possible. We only conduct analysis on the text content to prevent user re-identification.
    \item \textbf{Rate Limiting and Respect for Terms of Service:} 
    To minimize server load and comply with Reddit's Terms of Service, we throttled scraping operations. This ensured we did not disrupt service availability or degrade performance for other users.
\end{itemize}

Regarding deployment in production systems, we recognize that automated filtering decisions can impact user experience and perceived trust. Therefore, our proposed system architecture recommends transparent disclosure. Users must be informed when their contributions influence filtering decisions to maintain an expectation of reciprocity between the community and the platform's algorithms. We argue that this transparency mitigates the feeling of "surveillance capitalism" by aligning automated processes with user expectations regarding data usage.

While we did not seek Institutional Review Board (IRB) approval, as the data used in this work is publicly available, ethical reasoning was a key part of our design process. We prioritized minimizing harm over maximizing data utility where conflicts arose, ensuring that the benefits of improved filtering safety do not come at the cost of user privacy or trust.

\end{document}